\documentclass[%
reprint,
superscriptaddress,
amsmath,amssymb,
pra,
]{revtex4-2}

\usepackage{graphicx}
\usepackage{dcolumn}
\usepackage{bm}
\usepackage[colorlinks,linkcolor=blue, urlcolor=blue, anchorcolor=blue, citecolor=blue]{hyperref}
\usepackage{subfigure}
\usepackage{physics}

\usepackage{epstopdf}

\begin{document}
	
	\title{Relations between quantum metrology and criticality in general $su(1,1)$ systems}
	
	\author{Rui Zhang}
	\affiliation{Zhejiang Institute of Modern Physics, School of Physics, Zhejiang University, Hangzhou 310027, China}
	\author{Wenkui Ding}
	\affiliation{Key Laboratory of Optical Field Manipulation of Zhejiang Province and Department of Physics, Zhejiang Sci-Tech University, Hangzhou 310018, China}
	\author{Zhucheng Zhang}
	\email{zczhang@gscaep.ac.cn}
	\affiliation{Graduate School of China Academy of Engineering Physics, Beijing 100193, China}
	\author{Lei Shao}
	\affiliation{Graduate School of China Academy of Engineering Physics, Beijing 100193, China}
	\author{Yuyu Zhang}
	\affiliation{Department of Physics, Chongqing University, Chongqing 401330, China}
	\author{Xiaoguang Wang}
	\email{xgwang@zstu.edu.cn}
	\affiliation{Key Laboratory of Optical Field Manipulation of Zhejiang Province and Department of Physics, Zhejiang Sci-Tech University, Hangzhou 310018, China}
	\date{\today}

	\begin{abstract}
		There is a prevalent effort to achieve quantum-enhanced metrology using criticality. However, the extent to which estimation precision is enhanced through criticality still needs further exploration under the constraint of finite time resources. We clarify relations between quantum metrology and criticality through a unitary parametrization process with a Hamiltonian governed by $su(1,1)$ Lie algebra. We demonstrate that the determination of the generator in the parameterization can be treated as an extended brachistochrone problem. Furthermore, the dynamic quantum Fisher information about the parameter exhibits a power-law dependence on the evolution time as the system approaches its critical point. By investigating the dynamic sensing proposals of three quantum critical systems, we show that the asymptotic behavior of sensitivity is consistent with our predictions. Our theory provides a deep understanding on the interplay of quantum metrology and criticality, providing insights into the underlying connections that involve both quantum phenomena and classical problems.
		
	\end{abstract}
	\maketitle
	
	\section{Introduction}Quantum metrology and quantum sensing are promising quantum applications in quantum science and technology \cite{PhysRevLett.96.010401, RevModPhys.89.035002, toth2014quantum, giovannetti2011advances, PhysRevLett.121.020402,Braginsky_Khalili_Thorne_1992}.
	Fundamental quantum properties, such as coherence \cite{PhysRevLett.113.170401, Ares:21, PhysRevA.104.062608}  and entanglement \cite{PhysRevA.81.022108, PhysRevA.85.022322, PhysRevA.85.022321, PhysRevB.107.035123}, are leveraged to enhance the precision of parameter estimation.
	Specifically, entangled quantum probe states have been applied in quantum metrology to surpass the standard quantum limit, occasionally achieving the Heisenberg limit  \cite{PhysRevLett.98.090401, PhysRevLett.124.030501, giovannetti2004quantum, PhysRevLett.130.170801, quantum2013taddei}.
	However, preparing and maintaining highly entangled quantum states in many-body quantum systems faces challenges due to decoherence, limiting the achievable estimation precision in practical applications \cite{PhysRevLett.92.180403, Albarelli2018restoringheisenberg}.
	Recent exploration focuses on novel quantum properties, such as non-locality, correlations, and quantum discord, as avenues for achieving quantum-enhanced metrology and sensing \cite{khalid2018measurement,Modi2010QuantumCI,PhysRevLett.126.210506,PhysRevA.99.052318,Girolami_2015}.
	Criticality in quantum many-body systems is gaining theoretical and experimental attention for its potential in quantum-enhanced sensing \cite{Yang:19, PhysRevA.75.032109, PhysRevA.78.042105, PhysRevA.78.042106, PhysRevLett.126.200501, rams2018at, PRXQuantum.3.010354,yu2020experimental,PRXQuantum.3.010354,PhysRevA.105.042620}.
	Intuitively, the singularity at the critical point of a quantum phase transition indicates that a sight change in the parameter can notably alter the system's properties, implying the sensing utility.
	
	Basically, there exist two primary approaches to implement a criticality-enhanced sensing protocol.
	The first method utilizes the divergent fidelity susceptibility of the ground state near the critical point \cite{PhysRevE.76.022101,Ivanov_2020}.
	It introduces the parameter dependence into the ground state through an adiabatic quench of the system \cite{PhysRevLett.124.120504,PhysRevA.88.023803}.
	However, a notable practical challenge, termed ``critical slowing down'', is caused by the vanishing energy gap in many-body quantum critical systems, leading to the divergence of adiabatic evolution time \cite{rams2018at}.
	In contrast, the second method employs a sudden quench scheme, resembling conventional interferometric metrology \cite{DEMKOWICZDOBRZANSKI2015345}.
	Encoding of the parameter through Hamiltonian is set in the vicinity of the critical point, and the evolved quantum state is measured to estimate the parameter \cite{PhysRevLett.126.010502, Gietka2021adiabaticcritical, PhysRevA.93.022103,Skotiniotis_2015}.
	This dynamic framework avoids the critical slowing down problem, making it a more practically feasible option for experimental implementation.
	Interestingly, a recent study unifies these two methods and proves that the ultimate sensitivity of both is bounded by the same Heisenberg limit when the evolution time is explicitly considered, even though it exhibits super-Heisenberg scaling in certain many-body models within the first method \cite{rams2018at}.
	In addition, proposals involve harnessing criticality in quantum many-body systems to prepare highly entangled probe states for conventional interferometric sensing schemes \cite{PRXQuantum.2.030313, PhysRevLett.131.073201}.
	Besides, dissipative phase transition in open quantum systems has been explored to realize criticality-enhanced sensing \cite{PhysRevLett.120.150501, PhysRevA.97.013825, PhysRevX.13.031012,PhysRevA.96.013817,PhysRevE.89.022102}. 
	Furthermore, many research has concentrated on uncovering the origins of quantum enhancement in critical systems, exploring the novel inherent properties such as non-locality, symmetry breaking, gap closing, and long-range entanglement \cite{PhysRevLett.126.210506, braun2018quantum}.
	
	Apart from quantum many-body systems, the criticality in finite-component quantum systems has recently been investigated to realize quantum-enhanced sensing, encompassing both the ground state overlapping method and the sudden quench method \cite{PhysRevB.102.220302, gietka2022understanding}.
	The appeal of finite-component quantum critical systems lies in their experimental feasibility, characterized by a few degrees of freedom. The thermodynamic limit is easily achieved by adjusting specific system parameters. 
	A recent dynamic sensing procedure harnesses the criticality inherent in the finite-component quantum Rabi model (QRM), particularly near the critical point where dynamic quantum Fisher information (QFI) diverges \cite{PhysRevLett.126.010502}.
	Notably, this scheme circumvents the critical slowing down problem with non-stringent demands on probe state preparation. It enhances the appeal for real-world implementation in criticality-based metrology.
	However, in practical precision measurements, considering finite evolution time is essential for the sensitivity in the dynamic sensing scheme. Therefore, it is crucial to investigate the scaling behavior of the QFI with respect to the finite evolution time.
	
	In this paper, we analytically study the dynamic sensing schemes using a family of quantum critical systems, including the QRM, that can be described using the $su(1,1)$ Lie algebra.
	Through the derivation of an exact closed-form expression for the dynamic QFI, we reaffirmed that the sensitivity of parameter estimation is enhanced when the system is closed to its critical point. More importantly, we uncover a significant result: when the finite evolution time as an essential quantum resource is explicitly taken into account, the QFI exhibits a power-law dependence on the evolution time as the system approaches its critical point. Meanwhile, our study unveils a remarkable connection between the dynamics of the QFI and the classical brachistochrone problem. 
	This connection provides valuable insights into the characteristics and fundamental limitations of criticality-enhanced dynamic sensing schemes.
	
	\section{Quantum Fisher information for a general time-independent $su(1,1)$ Hamiltonian}The Hamiltonian governed by the $su(1,1)$ algebra takes the form of
	\begin{align}
		H_{\lambda} & =r_1 K_x +r_{2} K_y +r_{3} K_z 	=\boldsymbol{\mathrm{r}}\cdotp\boldsymbol{\mathrm{K}}, \label{eq:su11H}
	\end{align}
	where the vector $\boldsymbol{\mathrm{r}}=\boldsymbol{\mathrm{r}}(\lambda )$ remains constant over the evolution time $t$ within the parameter space of $\lambda$, and the operator $\boldsymbol{\mathrm{K}}=(K_x,K_y,K_z)$ is the generator of $su(1,1)$ algebra with components satisfying the commutation relations: $[K_x,K_y]=-iK_z$, $[K_y,K_z]=iK_x$, and $[K_z,K_x]\!=iK_y$. Differentiating the Hamiltonian with respect to $\lambda$ reads  $\partial_{\lambda}H_{\lambda}=\boldsymbol{\mathrm{\dot{r}}}\cdotp\boldsymbol{\mathrm{K}}$, where $\boldsymbol{\mathrm{\dot{r}}}=(\partial_{\lambda}r_1,\partial_{\lambda}r_2,\partial_{\lambda}r_3)$. The QFI associated with a state $\rho(\lambda)$ for the parameter $\lambda$ is defined as \cite{doi:10.1142/S0219749909004839,Helstrom,holevo2011probabilistic,liu2020quantum}: $F_\lambda:=\mathrm{Tr}(\rho L^2)$, where $\mathrm{Tr}$ represents trace and $L$ is the symmetric logarithmic derivative (SLD) operator, which is determined by $\partial_\lambda \rho=(L\rho+\rho L)/2$. 
	
\begin{figure}[t]
		\begin{centering}
			\includegraphics[scale=0.46]{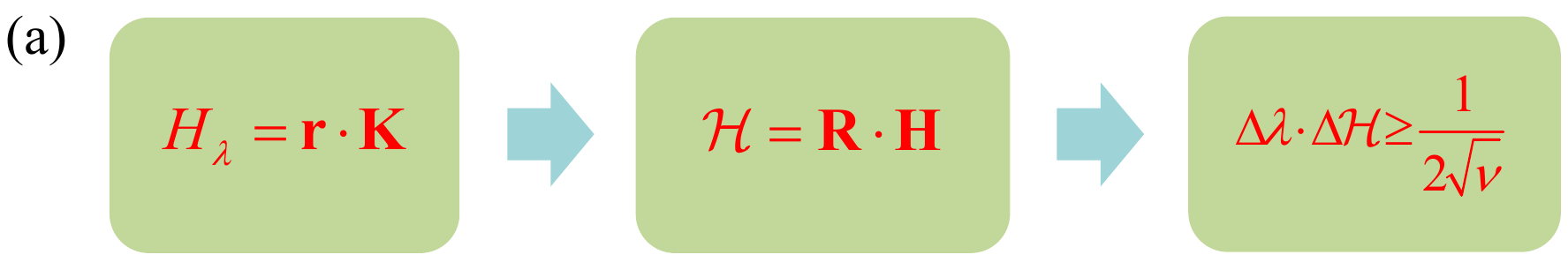}
			\includegraphics[scale=0.46]{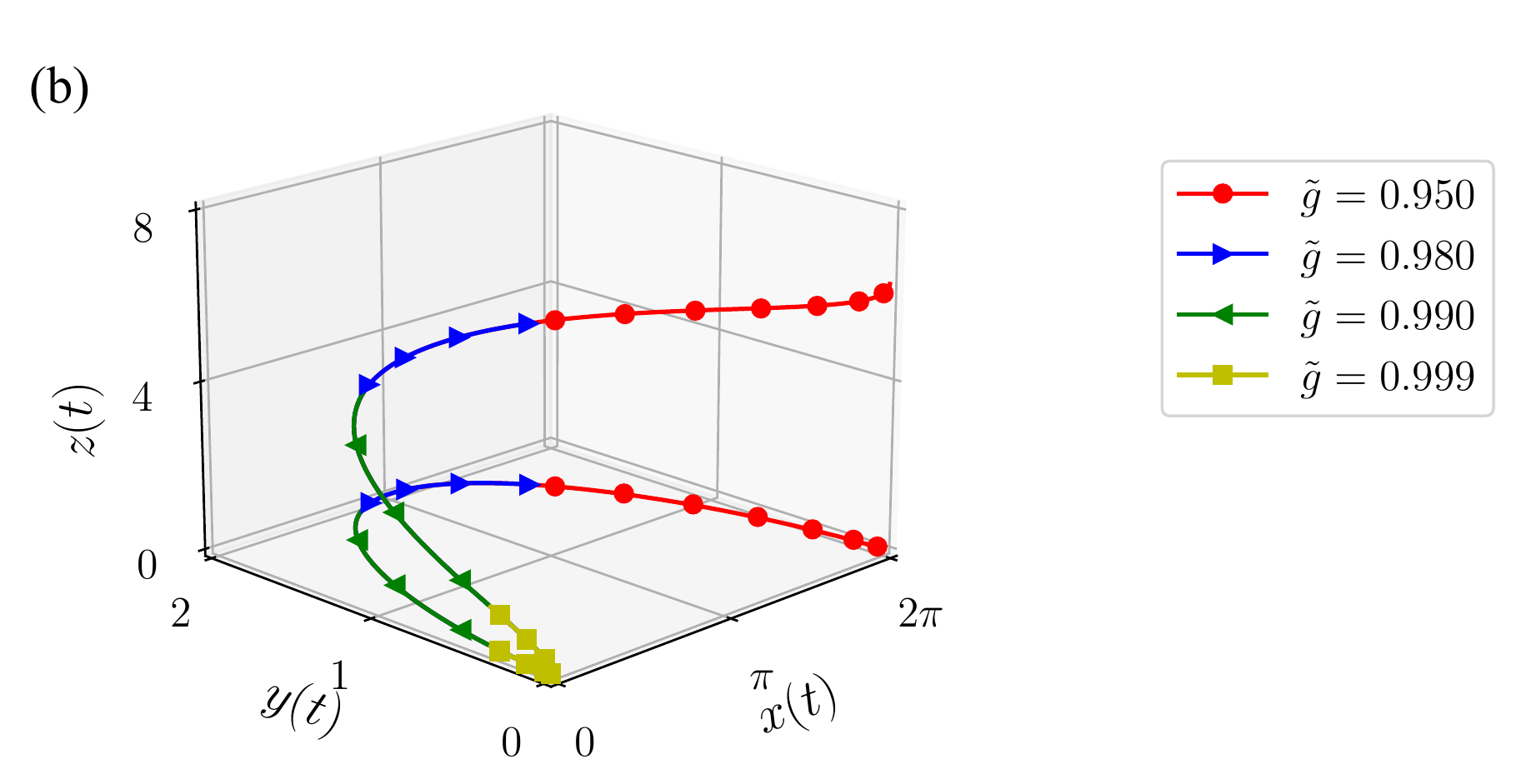}
		\end{centering}
		\caption{(a) Flow diagram depicting the connections among a Hamiltonian $H_\lambda$ characterized by a general form of $su(1,1)$ Lie algebra, the generator $\mathcal{H}$ of unitary parametrization, and a parameter-dependent uncertainty relationship. (b) Three-dimension graph and its projection onto the $x-y$ plane (the brachistochrone curve) for the Hermitian operator $\mathcal{H}$ in isotropic quantum Rabi model (i.e., $\zeta=1$) with various effective coupling strengths $(\tilde{g}=0.950, 0.980,0.990,0.999)$. The evolution time $t$ ranges from $0$ to $\pi/\omega\sqrt{1-\tilde{g}_0^2}$, where $\tilde{g}_0=0.950$ and the critical point occurs at $\tilde{g}=1$. }
		\label{fig1}
	\end{figure}

	Considering a general unitary parametrization process $U=e^{-iH_\lambda t}$ with a $\lambda$-independent initial state $|\psi_0\rangle$, the QFI can be expressed as the variance of a Hermitian operator $\mathcal{H}$ in the initial state, i.e., $F_\lambda=4\Delta^2[\mathcal{H}]_{|\psi_0\rangle}$, where  $\mathcal{H}:= i(\partial_{\lambda}U^{\dagger})U$ stands for the generator of parametrization process \cite{pang2017optimal,PhysRevA.90.022117,liu2015quantum}. The operator $\mathcal{H}$ has rich physical implications [see Fig.~\ref{fig1}(a)]. Firstly, it satisfies a Schr\"odinger-like equation with the evolution operator $U$, i.e., $i\partial _\lambda U^\dagger =\mathcal{H} U^\dagger$. Secondly, it connects the SLD through a relationship: $L=2iU[\mathcal{H}, \rho]U^{\dagger}$. Thirdly, it is linked to an uncertainty relationship with the parameter's precision $\Delta\lambda$, i.e., $\Delta \lambda \cdotp \Delta \mathcal{H} \geq \frac{1}{2\sqrt{\nu}}$,
	where $\nu$ represents the number of measurements, and $\Delta \mathcal{H} =\sqrt{\left\langle \mathcal{H}^2 \right\rangle -\left\langle \mathcal{H} \right\rangle^2 }$ denotes the standard deviation of $\mathcal{H}$. This uncertainty relationship places a constraint on the parameter's precision based on the generator $\mathcal{H}$.	
	
	For Hamiltonians that can be represented by Eq.~(\ref{eq:su11H}), the generator $\mathcal{H}$ can be derived as 
	\begin{align}
		\mathcal{H}&=-t h_z+i \sum_{n=0}^{\infty} \frac{(i t)^{2 n+2}}{(2 n+2) !} i^{2 n+1} r^{2 n} h_y\notag\\
		&+i \sum_{n=0}^{\infty} \frac{(i t)^{2 n+3}}{(2 n+3) !} i^{2 n+2} r^{2 n} h_x,
	\end{align}
	where $h_x = \mathbf{r}\boxtimes(\mathbf{r}\boxtimes\dot{\mathbf{r}})\cdotp\mathbf{K}$, $h_y = (\boldsymbol{\mathrm{r}}\boxtimes \boldsymbol{\mathrm{\dot{r}}}) \cdotp \boldsymbol{\mathrm{K}}$, and $h_z=\dot{\mathbf{r}}\cdotp\mathbf{K}$ are operators (see Appendix \ref{Appendix_A} for a derivation). $r=\sqrt{r_1^2+r_2^2-r_3^2}=\sqrt{\mathbf{r}\boxdot\mathbf{r}}$ is a number. Here, we propose a new operation rule for dot product $\boldsymbol{\mathrm{a}}\boxdot\boldsymbol{\mathrm{b}}$ and cross product $\boldsymbol{\mathrm{a}}\boxtimes\boldsymbol{\mathrm{b}}$ involving vectors $\boldsymbol{\mathrm{a}}$ and $\boldsymbol{\mathrm{b}}$. Given the singularity of $r$, the generator $\mathcal{H}$ is classified as follows: 
	
	(i) When $r^2_1+r^2_2>r^2_3$, $r=\mathbf{|r|}$, where $\mathbf{|r|} =\sqrt{|r_1^2+r_2^2-r_3^2|}$ is a real number. The generator can be simplified as
	\begin{align}
		\mathcal{H}	=\mathbf{R}\cdotp \mathbf{H},\label{r>0}
	\end{align}
	where $\mathbf{R}=(X(t),Y(t),Z(t))$ is a time-dependent coefficient vector, with components being 
	\begin{align}
		X(t) & = |\boldsymbol{\mathrm{r}}| t  -\sinh ( |\boldsymbol{\mathrm{r}}| t), \label{Eq:X} \\
		Y(t) & = 1 -\cosh  (|\boldsymbol{\mathrm{r}}| t), \label{Eq:Y}\\
		Z(t) & = |\boldsymbol{\mathrm{r}}| t.  \label{Eq:Z}
	\end{align}  
	And $\mathbf{H}=(h_x/\mathbf{|r|}^3,-h_y/\mathbf{|r|}^2, -h_z/\mathbf{|r|})$ is the time-independent vector. Specifically, the operator $h_x$ becomes $|\mathbf{r}|^3\partial_\lambda(\frac{\mathbf{r}}{|\mathbf{r}|})\cdotp\mathbf{K}$.
	
	(ii) Conversely, when $r^2_1+r^2_2<r^2_3$, $r=i\mathbf{|r|}$ becomes an imaginary number, and $h_x=-|\mathbf{r}|^3\partial_\lambda(\frac{\mathbf{r}}{|\mathbf{r}|})\cdotp\mathbf{K}$. Then the generator is represented as
	\begin{align}
		\mathcal{H^\prime}	=\mathbf{R}^\prime \cdotp \mathbf{H}^\prime.\label{r<0}
	\end{align}
	Here $\mathbf{H}^\prime=(-h_x/\mathbf{|r|}^3,h_y/\mathbf{|r|}^2, -h_z/\mathbf{|r|})$, and each component of $\mathbf{R}^\prime=\left( x(t), y(t), z(t) \right)$ is given by
	\begin{align}
		& x(t)= |\boldsymbol{\mathrm{r}}| t-\sin(|\boldsymbol{\mathrm{r}}| t), 	\label{eq:Brach4}\\
		& y(t)= 1-\cos(|\boldsymbol{\mathrm{r}}| t), \label{eq:Brach5}\\
		& z(t)= |\boldsymbol{\mathrm{r}}| t. \label{eq:Brach6}
	\end{align} 
	Notably, one can perceive that when $r$ is an imaginary value, the $x$ and $y$ components of $\mathbf{R}^\prime$ revert to the classical brachistochrone problem, which satisfies the cycloid equation, i.e., Eqs.~(\ref{eq:Brach4}) and (\ref{eq:Brach5}). 
	
	(iii) At the critical point of $r=0$, i.e., $r^2_1+r^2_2=r^2_3$, we have $h_x=0$. This implies that the generator contains only two non-zero terms, which is expressed as
	\begin{align}
		\mathcal{H}^{\prime \prime}=\mathbf{R}^{\prime \prime} \cdot \mathbf{H}^{\prime \prime} \label{r=0}
	\end{align}
	with $\mathbf{H}^{\prime \prime} =(0,h_y,-h_z)$  and $\mathbf{R}^{\prime \prime} =(0,\frac{t^2}{2},t)$. 
	
	Depending on the distinct values of $r$, we divide the generator $\mathcal{H}$ into three categories.
	Clearly, a discernible temporal dependency manifests within the generator.  
	In particular, in scenarios where $r\neq 0$, $\mathcal{H}$ can be interpreted as an extended brachistochrone problem on the $x-y$ plane, coupled with a linear dependence in the $z$ direction relative to the evolution time $t$. And the QFI can be decomposed into six parts (see Appendix \ref{Appendix_B}), characterized by $F_\lambda=4\sum_{s=1}^{6}F_s$. As the system approaches its critical point (i.e., $r\rightarrow 0$), we find that the QFI exhibits a power-law dependence on $t$, i.e.,
	\begin{equation}
		F_\lambda = \mathcal{A}_{xx}t^6 + \mathcal{A}_{yy}t^4 + \mathcal{A}_{zz}t^2 + 
		\mathcal{A}_{xy}t^5 + \mathcal{A}_{xz}t^4 + \mathcal{A}_{yz}t^3,
		\label{qfia}
	\end{equation}
	where $\mathcal{A}_{\alpha \beta}(\alpha,\beta=x,y,z)$ denote time-independent coefficients that vary with the initial probe state. Notably, for the large evolution time, the dominance of the sixth power term within the QFI becomes apparent, leading to an asymptotic expression $F_\lambda \simeq \mathcal{A}_{xx}t^6$. 
	
	In addition, under precise operation at the critical point (i.e., $r=0$), the expression of QFI is simplified as
	\begin{align}
		F_\lambda = \mathcal{B}_{yy}t^4 + \mathcal{B}_{yz}t^3 + \mathcal{B}_{zz}t^2,
		\label{qfib}
	\end{align}
	where $\mathcal{B}_{\alpha \beta}(\alpha, \beta=y,z)$ are time-independent coefficients. Overall, the dynamic QFI exhibits a power-law dependence on the evolution time at or near the critical point when the finite evolution time is explicitly taken into account. These remarkable results are illustrated and verified in subsequent examples. 
	
	\section{Quantum Rabi model}In this section, we explore the criticality-based dynamic sensing in QRM, whose Hamiltonian corresponds to the $su(1,1)$ algebra structure. The Hamiltonian of QRM reads ($\hbar =1$) \cite{PhysRevA.96.063821}
	\begin{align}
		H_{\mathrm{Rabi}} & =\omega a^{\dagger}a+\frac{\Omega}{2}\sigma_{z}+H_{c}, \\
		H_{c} & =g\left[(\sigma_{+}a+\sigma_{-}a^{\dagger})+\zeta(\sigma_{+}a^{\dagger}+\sigma_{-}a)\right].
	\end{align}
	Here, $a^{\dagger} (a)$ is the creation (annihilation) operator for the bosonic mode with frequency $\omega$; $\sigma_{k}(k=x,y,z)$ and $\sigma_{\pm}$ denote the Pauli operators and the raising/lowering operators, respectively, for the two-level system with transition frequency $\Omega$. The term $H_c$ arises from the coupling between the atom and bosonic field, characterized by the coupling strength $g$. The parameter $\zeta$ is a dimensionless coefficient that captures the ratio between the rotating and counter-rotating terms, and the Hamiltonian reduces to the isotropic QRM for $\zeta=1$.

	\subsection{Criticality in quantum Rabi model} \label{onesu11}
	Applying the Schrieffer-Wolff transformation on $H_{\mathrm{Rabi}}$, in the limit of $\Omega/\omega\rightarrow\infty$, one can derive a low-energy effective Hamiltonian for the normal phase in the spin-down subspace as \cite{PhysRevLett.119.220601,PhysRevLett.115.180404,PhysRev.149.491}
	\begin{align}
		H_{\mathrm{np}}^{(\downarrow)}=\frac{\omega}{2}(1-g_1^2)X_1^2+\frac{\omega}{2}(1-g_2^2)X_2^2,
		\label{sdh}
	\end{align}	
	where $X_1=(a+a^\dagger)/\sqrt{2}$ and $X_2=i(a^\dagger -a)/\sqrt{2}$ represent the dimensionless position and momentum operators, respectively (see Appendix \ref{Appendix_C} for a derivation). The parameters  $g_1=\tilde{g}(1+\zeta)/2$ and $g_2=\tilde{g}(1-\zeta)/2$ are defined in terms of the effective coupling strength $\tilde{g}=2g/\sqrt{\Omega\omega}$. The QRM undergoes a quantum phase transition at the critical point with $\tilde{g} = \tilde{g}_c=2/(1+|\zeta|)$ \cite{PhysRevLett.119.220601}. Under the one-mode bosonic realization of $su(1,1)$ algebra, i.e., $K_x=(a^2 + a^{\dagger2})/4, K_y=i(a^2 - a^{\dagger 2})/4$ and $K_z=(a^\dagger a+a a^\dagger)/4$, Eq.~(\ref{sdh}) becomes 
	\begin{align}
		H_{\mathrm{np}}^{(\downarrow)} =\omega(g_{2}^{2}-g_{1}^{2})K_x+\omega(2-g_{1}^{2}-g_{2}^{2})K_z.
		\label{suh}
	\end{align}
	According to Eq.~(\ref{eq:su11H}), we have the vector $\mathbf{r}=(\omega(g_{2}^{2}-g_{1}^{2}),0,\omega(2-g_{1}^{2}-g_{2}^{2}))$ in relation to the estimated parameter $\lambda=\tilde{g}$.  One can find that in the QRM with $\tilde{g}< \tilde{g}_c$, $r=i\mathbf{|r|}$, corresponding to case (ii); when $\tilde{g}=\tilde{g}_c$, $r=0$, corresponding to case (iii). 
	Specifically, for the isotropic case in the normal phase, we have $r=i2\omega \sqrt{1-\tilde{g}^2}$. Based on Eqs.~(\ref{eq:Brach4})-(\ref{eq:Brach6}), one can see that the period of the cycloid equation gives $\tau=\pi/\omega\sqrt{1-\tilde{g}^2}$, and it becomes infinite when the QRM operates at the phase transition point. As shown in Fig.~\ref{fig1}(b), the three-dimensional graph and its projection on the $x-y$ plane illustrate the impact of $\tilde{g}$ approaching the transition point on the cycloid period. As $\tilde{g}$ increases, it is clear that the brachistochrone curve becomes shorter and shorter, which indicates that more time is required to complete an entire periodogram. In other words, the evolution time tends to infinity when $\tilde{g}$ is close to $\tilde{g}_c$. 
	
	\begin{figure}
		\begin{centering}
			\includegraphics[scale=0.5]{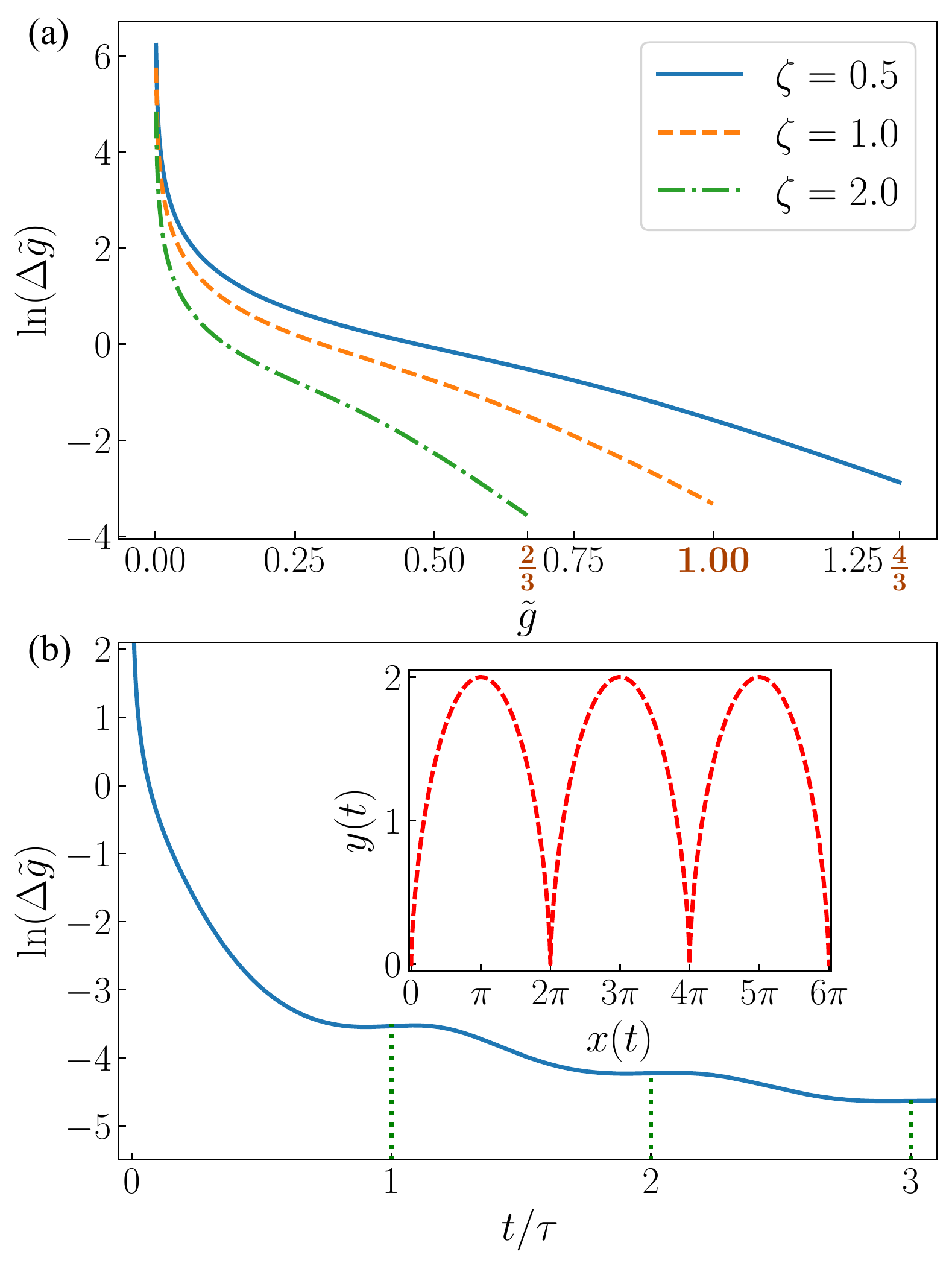}
		\end{centering}
		\caption{(a) Quantum Cram\'{e}r-Rao bounds of the effective coupling strength $\tilde{g}$ are plotted as a function of $\tilde{g}$ with different ratios $\zeta$ (i.e., $\zeta=0.5$, $\zeta=1.0$, $\zeta=2.0$), where the evolution time is fixed as $t=\pi/\omega$. (b) Quantum Cram\'{e}r-Rao bound of the isotropic QRM ($\zeta=1$) is plotted as a function of the evolution time $t$ with a fixed coupling strength $\tilde{g}_0=0.9$ and the corresponding evolution period $\tau=\pi/\omega\sqrt{1-\tilde{g}_0^2}$. The inset describes the cycloid equation in the $x-y$ plane. The initial state of the system is a product state between the two subsystems, that is, the two-level system is in its spin-down state $|\!\downarrow \rangle$ and the Bose field is $(|0\rangle+|1\rangle)/\sqrt{2}$.}\label{fig.2}
	\end{figure}
	
	In Fig.~\ref{fig.2}, we demonstrate the precision of the estimated parameter $\tilde{g}$, determined by the quantum Cram\'er-Rao bound \cite{cramer1999mathematical}: $\Delta \tilde{g} \geq 1/\sqrt{F_{\tilde{g}}}$, where $F_{\tilde{g}}$ denotes the QFI associated with $\tilde{g}$. Through a detailed analytical derivation (see Appendix \ref{Appendix_D}) , the uncertainty $\Delta \tilde{g}$ is found to be a function that depends on the evolution time $t$, the effective coupling strength $\tilde{g}$, and the ratio $\zeta$.
	We numerically simulate $\Delta \tilde{g}$ as the function of $\tilde{g}$ for various $\zeta$, while keeping the evolution time fixed at $t=\pi/\omega$ [see Fig.~\ref{fig.2}(a)]. The curves reveal that for different $\zeta$, $\Delta \tilde{g}$ decreases with the increase of $\tilde{g}$. Meanwhile, when $\tilde{g}$ approaches $\tilde{g}_c$ (i.e., $\frac{2}{3}$ for $\zeta=2.0$, $1$ for $\zeta=1.0$ and $\frac{4}{3}$ for $\zeta=0.5$), $\Delta \tilde{g}$ can obtain its minimum. 
	In addition, we analyze the relation of $\Delta \tilde{g}$ with the evolution time $t$ in the isotropic QRM under a given coupling strength $\tilde{g}_0=0.9$, as depicted in Fig.~\ref{fig.2}(b). 
	We observe a close connection between the evolution of $\Delta \tilde{g}$ and the brachistochrone problem. Specifically, $\Delta \tilde{g}$ periodically decreases with the evolution time, and the period $\tau$ is consistent with the cycloid equation (see the inset of the figure). Moreover, when the system approaches the phase transition point, the period is required to be infinite, reflecting the critical nature of the system.

	\subsection{Asymptotic behavior near the phase transition point} Now, we focus on the asymptotic behavior of QFI near the phase transition point (i.e., $\tilde{g}\rightarrow\tilde{g}_c$). We show that as the QRM tends to its phase transition point, taking into account the finite evolution time, an asymptotic expression of QFI is obtained as (see Appendix \ref{Appendix_D} for a derivation)
	\begin{equation}
		F_{\tilde{g}}\simeq \mathcal{A} t^6,
	\end{equation}
	where the factor $\mathcal{A}=320\zeta^4\omega^6/[9(1+\zeta )^6]$ is a function of the ratio $\zeta$.
	This expression implies that in the finite evolution time, the QFI near the phase transition point exhibits a proportionality to the sixth power of the evolution time and correlates with the ratio $\zeta$. 
	Additionally, as the QRM operates at its phase transition point (i.e., $\tilde{g}=\tilde{g}_c$), the expression of QFI in the finite evolution time is derived as
	\begin{equation}
		F_{\tilde{g}} = \mathcal{B}_1 t^4 + \mathcal{B}_2 t^2,
	\end{equation}
	where the factors $\mathcal{B}_1 = 16\zeta^2\omega^4/(1+\zeta)^2$ and $\mathcal{B}_2 = \omega^2[16\zeta^2 + (1+\zeta^2)^2]/(1+\zeta)^2$. Clearly, one can see that the above expressions of QFI verify our theory in Eqs.~(\ref{qfia}) and (\ref{qfib}).
	
\begin{figure}
		\begin{centering}
			\includegraphics[scale=0.5]{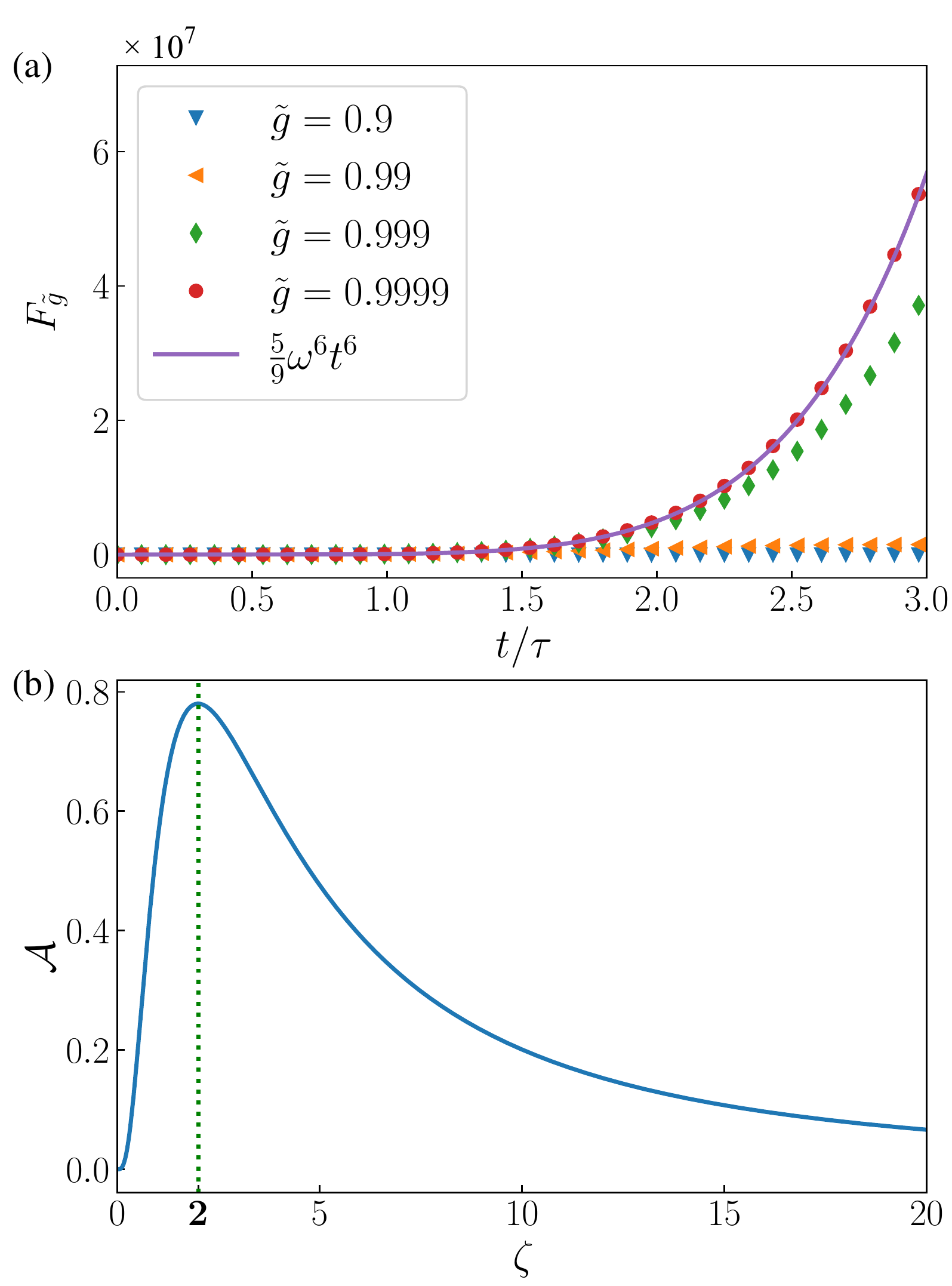}	
		\end{centering}
		\caption{(a) QFI $F_{\tilde{g}}$ of the isotropic QRM ($\zeta=1$) is plotted as a function the evolution time $t$ with different coupling strengths $\tilde{g}$ (i.e., $\tilde{g}=0.9,0.99,0.999,0.9999$). Here the value of the evolution period $\tau$ is fixed as $\pi/\omega\sqrt{1-\tilde{g}_0^2}$ with $\tilde{g}_0=0.9$. (b) Factor $\mathcal{A}$ in the asymptotic expression of the QFI is plotted as a function of the ratio $\zeta$.}
		\label{fig.3}
	\end{figure}

	To further illustrate the asymptotic behavior near the phase transition point, we numerically simulate the dynamic QFI of isotropic QRM, focusing on the QFI over the evolution time. As shown in Fig.~\ref{fig.3}(a), the evolution of QFI and its asymptotic expression ($F_{\tilde{g}}\simeq 5\omega^6 t^6/9$) are depicted. One can see that as $\tilde{g}\rightarrow1$, the dynamic QFI gradually fits the curve of the asymptotic expression. In addition, we also simulate the evolution of the factor $\mathcal{A}$ in the asymptotic expression with respect to the ratio $\zeta$, as presented in Fig.~\ref{fig.3}(b). It is apparent that the factor $\mathcal{A}$ reaches a maximum value at the ratio $\zeta=2$. This indicates that for the anisotropic QRM with $\zeta=2$, the QFI can attain its maximum when the system approaches the phase transition point. 
	
	\section{Extending the analysis to additional examples}In this section, to better illustrate our insights on the relation between quantum metrology and criticality, we extend the above results to additional two quantum critical systems. 
	The first one is the Lipkin-Meshkov-Glick model with the Hamiltonian \cite{PhysRevA.80.012318,LIPKIN1965188},
	\begin{align}
		H_{\mathrm{LMG}}=-\frac{1}{N}(S_x^2+\gamma S_y^2)-\eta S_z, \label{LMG}
	\end{align}
	where $S_\alpha=\sum_{i=1} ^N \sigma_{\alpha,i}/2  (\alpha=x,y,z)$ are the collective spin operators, $\sigma_{\alpha,i}$ denotes the Pauli matrices of the $i$th spin, and $N$ represents the total number of spins. The parameter $\gamma$ is the anisotropic parameter and $\eta$ describes the effective external field. By employing the Hosltein-Primakoff transformation and utilizing the one-mode realization of $su(1,1)$ algebra (see Appendix \ref{Appendix_E}), the Hamiltonian can be transformed into  $H_{\mathrm{LMG}}=(\gamma-1)K_x+(2\eta -\gamma-1)K_z$. This system experiences a spontaneous symmetry breaking at $\eta=1$, and when $\eta\rightarrow 1$, the asymptotic form of QFI regarding the parameter $\eta$ is derived as
	\begin{align}
		F_{\eta}\simeq\mathcal{A}_{\eta}t^6,
	\end{align}
	where the coefficient $\mathcal{A}_\eta=\frac{4(\gamma-1)^4}{9}\Delta^2[K_x-K_z]_{|\psi_0\rangle}$ is determined by the initial state $|\psi_0\rangle$.		
	
	The second example is the two-mode bosonic model with Pseudo-Anti-Parity-Time symmetry, described by the Hamiltonian \cite{PhysRevLett.128.173602},
	\begin{align}
		H_{\mathrm{APT}}=\delta(a^\dagger a+b^\dagger b)+i\kappa(a^\dagger b^\dagger-ab), \label{APT}
	\end{align}
	where $a(a^\dagger)$ and $b(b^\dagger )$ are the bosonic annihilation and creation operators in the two modes. The parameters $\delta $ and $\kappa$ correspond to the detuning and the coupling, respectively. With the two-mode representation of the $su(1,1)$ algebra, i.e., $K_x  =(a^\dagger b^\dagger + ab)/2$, $K_y=i(ab - a^\dagger b^\dagger)/2$, and $K_z =(a^\dagger a+b^\dagger b+1)/2$,
	the Hamiltonian can be rewritten as $H_{\mathrm{APT}}=2\delta K_z-2\kappa K_y$. This model undergoes a spontaneous symmetry breaking when $\kappa=\delta$. Choosing $\kappa$ as the parameter to be estimated, the asymptotic expression of QFI takes the form of  (see Appendix \ref{Appendix_E})
	\begin{align}
		F_{\kappa}\simeq \mathcal{A}_{\kappa}t^6,
	\end{align} 
	when $\kappa \rightarrow \delta$. Here the coefficient is expressed as $\mathcal{A}_{\kappa}=\frac{64}{9}\delta ^4 \Delta^2[K_y - K_z]_{|\psi_0 \rangle}$.
	
	The above examples once again verify the relation between quantum metrology and criticality. Indeed, for the criticality-enhanced dynamic sensing schemes in quantum systems described by the $su(1,1)$ algebra, the QFI about the parameter is proportional to the sixth power of the evolution time when the system tends to its critical point. 	
	
	\section{Discussion and conclusion}
	While the Hamiltonians under consideration adhere to the $su(1,1)$ algebra, our theoretical framework can also be applied to Hamiltonians governed by the $su(2)$ algebra  \cite{PhysRevA.92.012312,liang2018quantum}.
	In these Hamiltonians, the connection between the dynamics of QFI and the brachistochrone problem, as well as the power-law dependency of QFI on the evolution time at the critical point, can still be revealed. 
	On the other hand, the QFI described by the function $t^6$ is divergent when the evolution time tends to infinity. However, in the implementation of the dynamic quantum sensing, the finite evolution time should be properly considered as a quantum resource. 
	
	In summary, we have investigated the sensitivity limits of criticality-enhanced dynamic sensing schemes. By incorporating the evolution time as a quantum resource, we have revealed that the QFI near the critical point follows a power-law dependence on the evolution time. Notably, our results also unveil a close connection between the dynamics of QFI and the classical brachistochrone problem, offering immediate insights into the time dependence.  
	
	\begin{acknowledgments}
		This work was supported by the National Natural Science Foundation of China (Grants No.~11935012, No.~	
		12347123 and No.~12347127) and the Hangzhou Joint Fund of the Zhejiang Provincial Natural Science Foundation of China (Grant No. LHZSD24A050001).
	\end{acknowledgments}
	
	\appendix
	\section{Generator in a general $su(1,1)$ parametrization process}\label{Appendix_A}

	Here, we give a detailed derivation process for the expression of $\mathcal{H}$ based on the $su(1,1)$ Hamiltonian $H_\lambda$ in Eq.~(\ref{eq:su11H}).
	The generator $\mathcal{H}=i(\partial_{\lambda}U^{\dagger})U$ is represented in a Taylor expansion with the unitary operator $U=e^{-iH_\lambda t}$ as \cite{liu2015quantum}
	\begin{align}
		\mathcal{H} & =i \sum_{n=0}^{\infty} \frac{(i t)^{n+1}}{(n+1) !} H_\lambda^{\times n}\left(\partial_\lambda H_\lambda\right) \notag\\
		& =-t H_\lambda^{\times 0}\left(\partial_\lambda H_\lambda\right)+i \sum_{n=0}^{\infty} \frac{(i t)^{2 n+2}}{(2 n+2) !} H_\lambda^{\times(2 n+1)}\left(\partial_\lambda H_\lambda\right) \notag\\
		&+i \sum_{n=0}^{\infty} \frac{(i t)^{2 n+3}}{(2 n+3) !} H_\lambda^{\times(2 n+2)}\left(\partial_\lambda H_\lambda\right),
		\label{genH}
	\end{align}
	where the superoperator $H_\lambda ^{\times n}(\cdot)=[H_\lambda,\cdots, [H_{\lambda},\cdot]]$ denotes a $n$th-order nested commutator operation. Due to the commutation relations between the generators of $su(1,1)$ algebra, one can find that there are two terms with the forms $(a_1b_1+a_2b_2-a_3b_3)$ and $(a_2b_3-a_3b_2, a_3b_1-a_1b_3, a_2b_1-a_1b_2)$, where $a_j(j=1,2,3)$ and $b_j(j=1,2,3)$ are components of vectors $\boldsymbol{\mathrm{a}}$ and $\boldsymbol{\mathrm{b}}$, respectively. For convenience, we set $(a_1b_1+a_2b_2-a_3b_3)=\boldsymbol{\mathrm{a}}\boxdot\boldsymbol{\mathrm{b}}$ and $(a_2b_3-a_3b_2, a_3b_1-a_1b_3, a_2b_1-a_1b_2)=\boldsymbol{\mathrm{a}}\boxtimes\boldsymbol{\mathrm{b}}$, which are different from the traditional dot product $\boldsymbol{\mathrm{a}}\cdotp\boldsymbol{\mathrm{b}}$ and cross product $\boldsymbol{\mathrm{a}}\times\boldsymbol{\mathrm{b}}$, respectively. Under the new vector operation rule, we obtain a new commutation relation, i.e.,
	\begin{align}
		[\boldsymbol{\mathrm{a}}\cdot\boldsymbol{\mathrm{K}},\boldsymbol{\mathrm{b}}\cdot\boldsymbol{\mathrm{K}}] =i(\boldsymbol{\mathrm{a}}\boxtimes\boldsymbol{\mathrm{b}})\cdot\boldsymbol{\mathrm{K}},
	\end{align}
	and a new Lagrange formula,
	\begin{align}
		\boldsymbol{\mathrm{a}}\boxtimes(\boldsymbol{\mathrm{b}}\boxtimes\boldsymbol{\mathrm{c}})=-(\boldsymbol{\mathrm{a}}\boxdot\boldsymbol{\mathrm{c}})\boldsymbol{\mathrm{b}} +(\boldsymbol{\mathrm{a}}\boxdot\boldsymbol{\mathrm{b}})\boldsymbol{\mathrm{c}}.
	\end{align}
	Then with the above new formulas, we have
	\begin{align}
		H_\lambda^{\times} (\partial_{\lambda}H_\lambda)&=[\mathbf{r} \cdot \mathbf{K}, \dot{\mathbf{r}} \cdot \mathbf{K}]=i(\mathbf{r}\boxtimes\mathbf{\dot{r}})\cdotp\mathbf{K},\label{h1}
	\end{align}
	and 
	\begin{align}
		H_\lambda^{\times 2} (\partial_{\lambda}H_\lambda)&= [\mathbf{r} \cdot \mathbf{K}, i(\mathbf{r}\boxtimes\mathbf{\dot{r}})\cdotp\mathbf{K}]=-\mathbf{r}\boxtimes(\mathbf{r}\boxtimes\dot{\mathbf{r}})\cdotp\mathbf{K}\notag\\
		&      =r(\dot{r}\mathbf{r}-r\dot{\mathbf{r}})\cdotp\mathbf{K},\label{h2}
	\end{align}
	where $r=\sqrt{r_1^2+r_2^2-r_3^2}=\sqrt{\mathbf{r}\boxdot\mathbf{r}}$ and we have employed the formula $\mathbf{r}\boxdot\mathbf{\dot{r}}=\frac{1}{2}\frac{d(r^2)}{d\lambda}=r\dot{r}$.
	
	Without loss of generality, denoting $h_x\! =\!-H_\lambda^{\times 2} (\partial_{\lambda}H_\lambda)$, $\!h_y\!=-iH_\lambda^{\times} (\partial_{\lambda}H_\lambda)$ and $h_z=H_\lambda^{\times 0} (\partial_{\lambda}H_\lambda)$, it is proven that $h_x$ and $ h_y$ are eigenoperators of the superoperator, i.e.,
	\begin{align}
		H_\lambda^{\times 2n+1} (\partial_{\lambda}H_\lambda) = i^{2n+1} r^{2n} h_y, \notag\\
		H_\lambda^{\times 2n+2} (\partial_{\lambda}H_\lambda) = i^{2n+2} r^{2n}h_x. \label{Ht2n3}
	\end{align}
	Then the generator $\mathcal{H}$ in Eq.~(\ref{genH}) can be rewritten as
	\begin{align}
		\mathcal{H}=&-t h_z+i \sum_{n=0}^{\infty} \frac{(i t)^{2 n+2}}{(2 n+2) !} i^{2 n+1} r^{2 n} h_y \notag \\
		&+i \sum_{n=0}^{\infty} \frac{(i t)^{2 n+3}}{(2 n+3) !} i^{2 n+2} r^{2 n} h_x.
	\end{align}
	
	Given the particularity of $r$, the generator $\mathcal{H}$ should be classified and discussed. First, for $r\neq 0$, we have 
	\begin{align}
		\mathcal{H} & =-t h_z+\sum_{n=0}^{\infty} \frac{(r t)^{2 n+2}}{(2 n+2) !} \frac{h_y}{r^2}-\sum_{n=0}^{\infty} \frac{(r t)^{2 n+3}}{(2 n+3) !} \frac{h_x}{r^3} \notag\\
		& =-t h_z-\frac{1-\cosh (r t)}{r^2} h_y+\frac{r t-\sinh (r t)}{r^3} h_x,
	\end{align}
	and the operator $h_x$ can be simplified to $r^3 \partial_\lambda (\frac{\mathbf{r}}{r})\cdotp\mathbf{K}$ by employing the relationship $\partial_\lambda (\frac{\mathbf{r}}{r})=\frac{r\mathbf{\dot{r}}-\dot{r}\mathbf{r}}{r^2}$. 
	Then, based on whether $r$ is a real number or an imaginary number, the above equation can be further divided into two cases, which correspond to the cases (i) and (ii). In addition, for $r=0$ (i.e., $r^2_1+r^2_2=r^2_3$), $h_x=0$, so we have
	\begin{align}
		\mathcal{H} =-t h_z+\frac{(i t)^2}{2!} i^2 h_y 
		=-t h_z+\frac{t^2}{2} h_y,
	\end{align}
	which corresponds to the case (iii).
	
	\section{Expression for QFI in quantum systems governed by $su(1,1)$ algebra}\label{Appendix_B}
	Upon determining the distinct value of $r$, one can obtain the corresponding generator $\mathcal{H}$ using $su(1,1)$ parametrization. Subsequently, the QFI  regarding the estimated parameter $\lambda$ is given by
	\begin{align}
		F_{\lambda} =4\Delta ^2 [\mathcal{H}]_{|\psi_0\rangle}.\label{Fgb1}
	\end{align}
	Whether $r$ is real or imaginary, the identical asymptotic equation can be obtained. 
		Taking the case (ii) as an example, the QFI can be decomposed into six parts, i.e., $F_{\lambda }=4\sum_{s=1}^{6}F_s$. The specific forms for each components are provided below 
	\begin{align}
		F_{1} & =z^{2}\frac{\Delta^{2}[h_z]_{|\psi_0\rangle}}{\mathbf{|r|}^2}, \quad	
		F_{2}=-2yz\frac{\mathrm{Cov}[h_y,h_z]_{|\psi_0\rangle}}{\mathbf{|r|}^3}, \nonumber \\
		F_{3} & =2xz\frac{\mathrm{Cov}[h_x,h_z]_{|\psi_0\rangle}}{\mathbf{|r|}^4}, \quad
		F_{4}=y^{2}\frac{\Delta^{2}[h_y]_{|\psi_0\rangle}}{\mathbf{|r|}^4}, \nonumber \\
		F_{5} & =-2xy\frac{\mathrm{Cov}[h_x,h_y]_{|\psi_0\rangle}}{\mathbf{|r|}^5}, \quad
		F_{6}=x^{2}\frac{\Delta^{2}[h_x]_{|\psi_0\rangle}}{\mathbf{|r|}^6},
		\label{F_k2}
	\end{align}
	where $\Delta^{2}[o]_{|\psi_0\rangle}=\left\langle o^{2}\right\rangle -\left\langle o\right\rangle ^{2}$
	and $\mathrm{Cov}[o_{m},o_{n}]_{|\psi_0\rangle}=\frac{1}{2}\left\langle \{o_{m},o_{n}\}\right\rangle -\left\langle o_{m}\right\rangle \left\langle o_{n}\right\rangle $ represent the variance and covariance of the operators $h_{x,y,z}$ in the initial state $|\psi_0\rangle$, respectively. When the finite evolution time are taken into account, in the limit of $r\rightarrow0$, we have
	\begin{align}
		F_{\lambda}=\mathcal{A}_{xx}t^6+\mathcal{A}_{yy}t^4+\mathcal{A}_{zz}t^2+\mathcal{A}_{xy}t^5+\mathcal{A}_{xz}t^4+\mathcal{A}_{yz}t^3, \label{Flambda}
	\end{align}
	where $\mathcal{A}_{xx}\!\!=\!\!\frac{1}{9}\underset{|\mathbf{r}|\rightarrow0}{\mathrm{lim}}\Delta^{2}[h_x]_{|\psi_0\rangle}$, $\mathcal{A}_{yy}\!\!=\!\!\!\underset{|\mathbf{r}|\rightarrow0}{\mathrm{lim}}\Delta^{2}[h_y]_{|\psi_0\rangle}$, $\mathcal{A}_{zz}\!\!=\!\!\!\underset{|\mathbf{r}|\rightarrow0}{\mathrm{lim}}4\Delta^{2}[h_z]_{|\psi_0\rangle}$, $\mathcal{A}_{xy}=-\frac{2}{3}\underset{|\mathbf{r}|\rightarrow0}{\mathrm{lim}}\mathrm{Cov}[h_x,h_y]_{|\psi_0\rangle}$, $\mathcal{A}_{xz}\!\!\!=\!\!\!\frac{4}{3}\underset{|\mathbf{r}|\rightarrow0}{\mathrm{lim}}\mathrm{Cov}[h_x,h_z]_{|\psi_0\rangle}\!$ and $\!\mathcal{A}_{yz}=-\underset{|\mathbf{r}|\rightarrow0}{\mathrm{lim}}4\mathrm{Cov}[h_y,h_z]_{|\psi_0\rangle}$.
	
	In addition, for the critical point of $r=0$, the QFI can be derived as
	\begin{align}
		F_\lambda = \mathcal{B}_{yy}t^4 + \mathcal{B}_{yz}t^3 + \mathcal{B}_{zz}t^2,
		\label{qfib2}
	\end{align}
	where $\mathcal{B}_{yy}=\Delta^{2}[h_y]_{|\psi_0\rangle}$, $\mathcal{B}_{yz}=-4\mathrm{Cov}[h_y,h_z]_{|\psi_0\rangle}$ and $\mathcal{B}_{zz}=4\Delta^{2}[h_z]_{|\psi_0\rangle}$.
	
	\section{Effective Hamiltonian of QRM in the form of $su(1,1)$ algebra}\label{Appendix_C}
	
	For the completeness of the content, we present the derivation process of the low-energy effective Hamiltonian for QRM, and then obtain the Hamiltonian in the form of $su(1,1)$ algebra. 
	
	The Hamiltonian of QRM can be expressed in terms of dimensionless position and momentum operators, $X_1=(a^{\dagger}+a)/\sqrt{2}$ and $X_2=i(a^{\dagger}-a)/\sqrt{2}$, with the form of
	\begin{align}
		H_{\mathrm{Rabi}} &=H_0+H_c,\label{H_initial} \notag\\
		H_0 &= \frac{\omega }{2}(X_2^2+X_1^2)+\frac{\Omega}{2}\sigma _z,\notag\\
		H_c &=g [(1+\zeta )\sigma_x X_1-(1-\zeta  )\sigma _y X_2]/\sqrt{2},
	\end{align}
	where $H_0$ is the unperturbed part and $H_{c}$ describes the coupling between the spin-up subspace $(\uparrow)$ and spin-down subspace $(\downarrow)$. The derivation of the decoupled Hamiltonian involves employing the Schrieffer-Wolff (SW) unitary transformation on $H_{\mathrm{Rabi}}$ in the limit of $\Omega/ \omega \rightarrow \infty$ \cite{PhysRev.149.491,PhysRevA.95.013819}. The transformation process unfolds as follows:
	
	Employing a unitary transformation with an operator $U_{\mathrm{SW}}=\exp(S)$ on $H_{\mathrm{Rabi}}$, where $S$ represents an anti-Hermitian operator, one can obtain the transformed Hamiltonian as
	\begin{align}
		H_{\mathrm{eff}}=U_{\mathrm{SW}}^\dagger  H_{\mathrm{Rabi}} U_{\mathrm{SW}}.
	\end{align}
	The operator $S$ can be approximated as $S\simeq S_1+S_3$, where the operator $S_1$ and $S_3$ satisfy specific conditions, i.e., $[H_{0},S_{1}] =-H_{c}$ and $[H_{0},S_{3}] =-\frac{1}{3}[[H_{c},S_{1}],S_{1}]$. In the limit of $\Omega/ \omega \rightarrow \infty$, neglecting the high-order terms, one can derive the resulting second-order effective Hamiltonian as 
	\begin{align}
		H_{\mathrm{eff}}  \simeq H_0 +\frac{1}{2}[H_c,S_1],
	\end{align}
	with the operator $S_1$ in the following form
	\begin{align}
		S_{1} & =-i\tilde{g}\sqrt{\frac{\omega }{8\Omega }}[(1-\zeta )\sigma_{x}X_2+ (1+\zeta )\sigma_{y}X_1].
	\end{align}
	
	In the spin-down ($\downarrow$) subspace, the effective Hamiltonian takes the form of
	\begin{align}
		H_{\mathrm{np}}^{(\downarrow )} =\frac{\omega}{2}(1-g_{1}^{2})X_1^{2}+\frac{\omega}{2}(1-g_{2}^{2})X_2^{2}. \label{H_eff} 
	\end{align}
	Utilizing the one-mode bosonic realization of $su(1,1)$ algebra, 
	we have $X_1^2=2(K_x+K_z)$, and $X_2^2=2(K_z-K_x)$. Then the effective Hamiltonian can be reformulated as 
	\begin{align}
		H_{\mathrm{np}}^{(\downarrow)}=r_1K_x + r_2K_y + r_3K_z, \label{Hnpr}
	\end{align}
	where $r_1 = \omega(g_{2}^{2}-g_{1}^{2})$, $r_2=0$ and $r_3=\omega(2-g_{1}^{2}-g_{2}^{2})$.	Note that the constant terms in the Hamiltonian have no impact on the QFI, so for simplicity, we ignore them in our expressions. 
	
	\section{Asymptotic expression of QFI in QRM}\label{Appendix_D}
	\begin{figure}[b]
		\centering\includegraphics[scale=0.6]{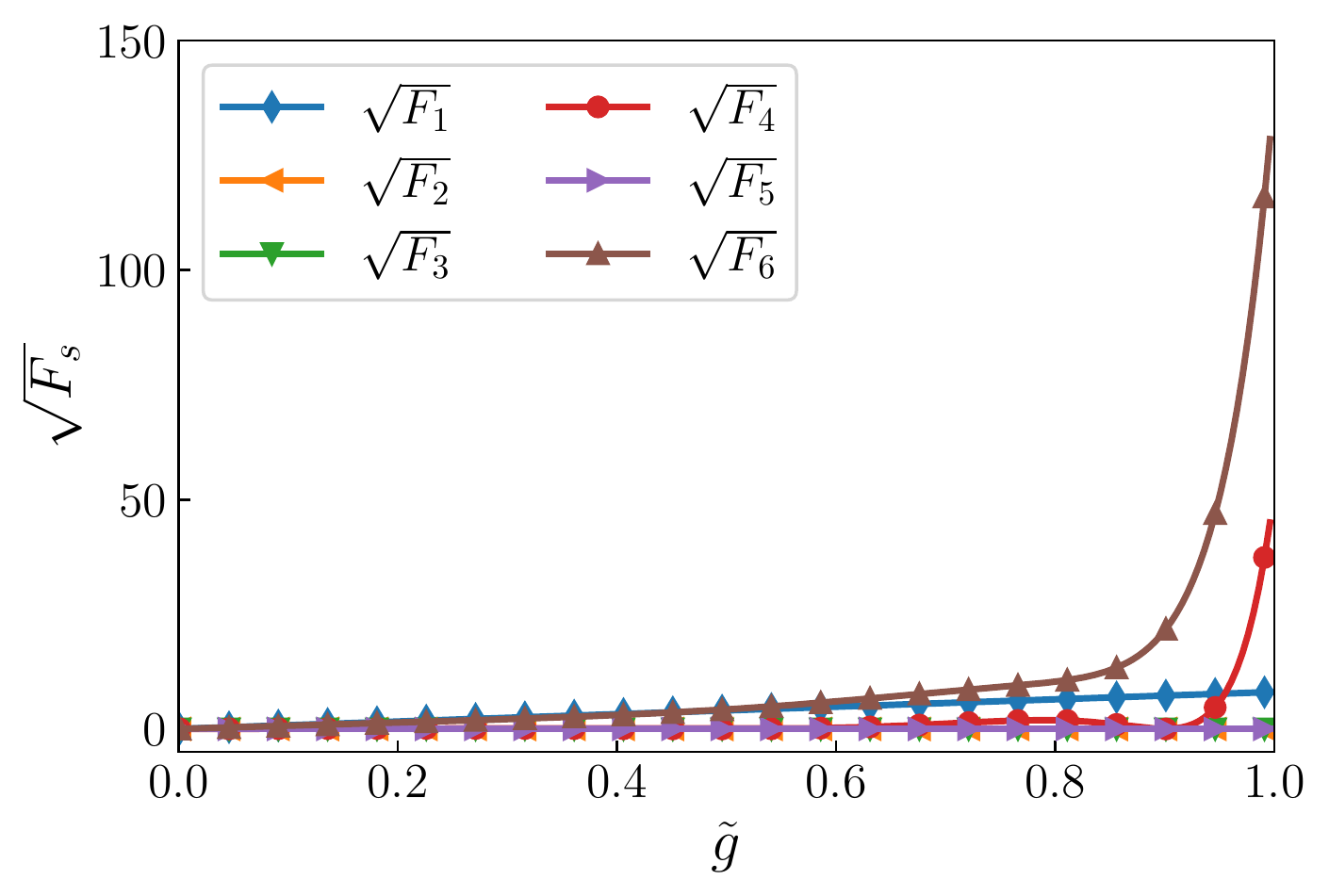}
		\caption{Variation of the six parts of QFI ($F_s$) with respect to the effective coupling strength $\tilde{g}$, where the evolution time is fixed as $t=\pi/\omega\sqrt{1-\tilde{g}_0^2}$ with $\tilde{g}_0=0.9$, and the ratio $\zeta=1$. The initial state is consistent with the one presented in the above figure.}\label{A1}
	\end{figure}
\begin{figure}[h]
		\begin{centering}
			\includegraphics[scale=0.62]{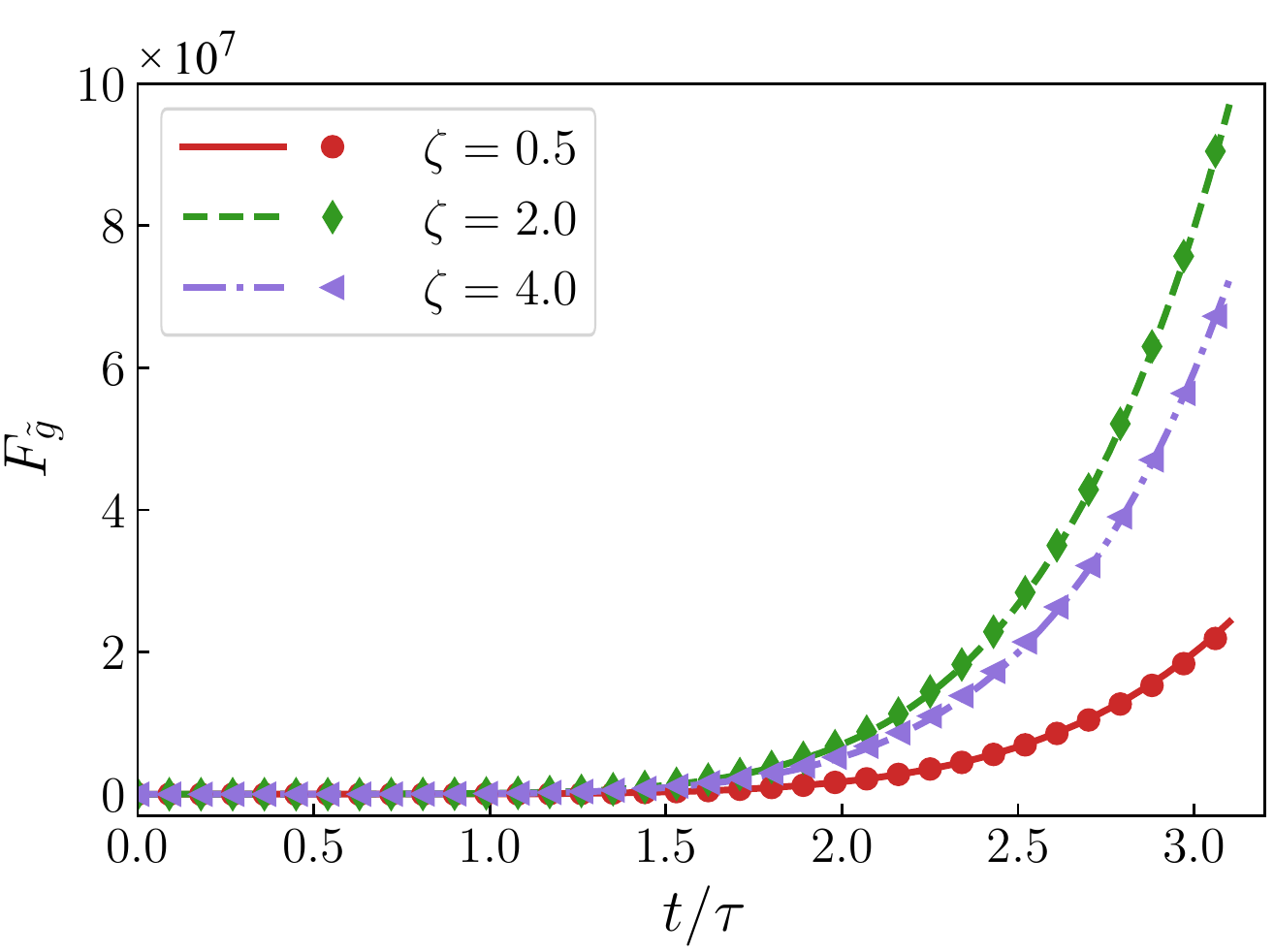}
		\end{centering}
		\caption{Evolution contrast between the QFI $ F_{\tilde{g}}$ and its asymptotic curve when the system approaches the phase transition point $\tilde{g}_c$ for different ratios $\zeta$ (i.e., $\zeta=0.5$, $\zeta=2.0$, $\zeta=4.0$).}\label{A2}
	\end{figure}

	Below, we derive the asymptotic expression of QFI in the normal phase of QRM. Referring to Eq.~(\ref{Hnpr}), for the QRM with $\tilde{g} < \tilde{g}_c$, we have
	\begin{align}
		h_x&=(r_1 \dot{r}_3-r_3 \dot{r}_1 )(r_3 K_x + r_1 K_z), \notag\\
		h_y&=(r_3 \dot{r}_1 -r_1 \dot{r}_3)K_y,\notag\\
		h_z&=\dot{r}_1 K_x + \dot{r}_3 K_z.
	\end{align}
	And based on Eq.~(\ref{Fgb1}), the exact expression of QFI is obtained.	We analyze the evolution of the QFI with the estimated parameter $\tilde{g}$. As shown in Fig.~\ref{A1}, we numerically simulate the change of six parts of the QFI, and find that a significant increase in the contribution of the sixth part ($F_6$) as the system tends to the phase transition point infinitely. Thus, for the QRM with $\tilde{g} \rightarrow \tilde{g}_c$, one can find that the QFI can be approximated as
	\begin{align}
		F_{\tilde{g}}&\simeq 4F_6 = \frac{4x^{2}}{|\mathbf{r}|^{6}}\frac{4^5\omega^6 \zeta^4}{(1+\zeta)^6}\Delta^2[K_x-K_z]_{|\psi_0\rangle} \simeq \mathcal{A} t^6,
	\end{align}
	where the factor is $\mathcal{A}=\frac{320\zeta^{4}\omega^6 }{9(1+\zeta)^{6}}$. The factor $\mathcal{A}$ is influenced by the ratio $\zeta$.	In the isotropic quantum Rabi model, i.e., $\zeta=1$, the above expression for $F_{\tilde{g}}$ becomes 	
	\begin{equation}
		F_{\tilde{g}}\simeq \frac{5}{9}\omega^6 t^6. \label{F6}
	\end{equation}
	In Fig.~\ref{fig.3}, we have confirmed the asymptotic behavior of QFI in accordance with  Eq.~(\ref{F6}) under the isotropic condition. Here, we compare the results in different parameters $\zeta$ close to the phase transition point, as illustrated in Fig.~\ref{A2}. From the curves, one can see that the asymptotic curves can fit the QFI very well.
	
	For the QRM at the critical point $\tilde{g} = \tilde{g}_c$, except for $h_x=0$, the expressions for $h_y$ and $h_z$ remain consistent with the formulations previously presented in the normal phase. Employing the expression of the QFI in Eq.~(\ref{qfib2}), we obtain
	\begin{equation}
		F_{\tilde{g}} = \mathcal{B}_1 t^4 + \mathcal{B}_2 t^2,
	\end{equation}
	where the factors $\mathcal{B}_1 = 16\zeta^2\omega^4/(1+\zeta)^2$ and $\mathcal{B}_2 = \omega^2[16\zeta^2 + (1+\zeta^2)^2]/(1+\zeta)^2$. Specifically, for the isotropic QRM, the QFI simplifies to $F_{\tilde{g}}=5(\omega t)^2 +4(\omega t)^4$.

	\section{Two other models governed by the $su(1,1)$ algebra} \label{Appendix_E}
	
	In this Appendix, we derive the asymptotic expression of two additional models governed by the $su(1,1)$ algebra. 
	
	The first example is the Lipkin-Meshkov-Glick model, a basic quantum model of interacting spins characterized by the Hamiltonian in Eq.~(\ref{LMG}). In the thermodynamic limit, the effective Hamiltonian for the model can be derived through the Holstein-Primakoff transformation, which leads to the following relations \cite{LIPKIN1965188}
	\begin{align}
		S_{+} =\sqrt{N}\sqrt{1-\frac{a^{\dagger}a}{N}}&a, \quad
		S_-  =\sqrt{N}a^{\dagger}\sqrt{1-\frac{a^{\dagger}a}{N}}, \notag \\
		S_{z} & =\frac{N}{2}-a^{\dagger}a.
	\end{align}
	In the low excitation regime, characterized by $\left\langle a^{\dagger}a\right\rangle /N\ll1$, the approximation $\sqrt{1-a^{\dagger}a/N}\approx1$ holds. Then we have
	\begin{align}
		H_{\mathrm{LMG
		}}\approx\left[\gamma\left(a^{\dagger}-a\right)^2-\left(a+a^{\dagger}\right)^2\right] / 4+\eta a^{\dagger} a,
	\end{align}
	which is further written in the form of  the one-mode bosonic realization in $su(1,1)$ algebra, i.e.,
	\begin{align}
		H_{\mathrm{LMG}} = (\gamma-1) K_{x} + (2\eta-\gamma-1)K_{z}.
	\end{align}
	Note that in the above derivation process, the constant terms have been omitted. By comparing with Eq.~(\ref{eq:su11H}), one can find that the components of the vector $\mathbf{r}$ are: $r_1=\gamma-1$, $r_2=0$ and $r_3=2\eta-\gamma-1$, so that $r=2\sqrt{(\eta-\gamma)(1-\eta)}$. 
	Choosing $\eta$ as the parameter to be estimated, for $\eta\neq 1$, we have
	\begin{align}
		h_x&=2r_1(r_3K_{x}+r_1K_{z}), \notag\\
		h_y&=-2r_1K_{y},\notag\\
		h_z&=2K_{z},
	\end{align}
	then the asymptotic expression of QFI about the parameter $\eta$ can be expressed as
	\begin{align}
		F_{\eta}\simeq \frac{4(\gamma-1)^4}{9}t^6\Delta^2[K_x-K_z]_{|\psi_0\rangle}, \quad \eta\rightarrow 1.
	\end{align}	
	For the critical point of $\eta=1$, we have $h_x=0$, and the QFI becomes
	\begin{align}
		F_\eta=&4(\gamma-1)^2 \Delta^2[K_y]_{|\psi_0\rangle} t^4 +16\Delta^2[K_z]_{|\psi_0\rangle} t^2 \notag\\
		&+16 (\gamma-1)\mathrm{Cov}[K_y,K_z]_{|\psi_0\rangle} t^3.
	\end{align}
	
	The second example is the two-mode bosonic model with pseudo-Anti-Parity-Time symmetry. Under the two-mode representation of the $su(1, 1)$ algebra with generators, the Hamiltonian in Eq.~(\ref{APT}) can be written as
	\begin{align}
		H_{\mathrm{APT}}=2\delta K_z-2\kappa K_y.
	\end{align}
	Similarly, by comparing with Eq.~(\ref{eq:su11H}), one can see that the components of the vector $\mathbf{r}$ are: $r_1=0$, $r_2=-2\kappa$ and $r_3=2\delta$ and $r=2\sqrt{(\kappa + \delta)(\kappa - \delta)}$. Choosing $\kappa$ as the parameter to be estimated, for $\kappa \neq \delta$, we have
	\begin{align}
		h_x&= -r_3 \dot{r}_2 (r_3 K_y+r_2K_z), \notag\\
		h_y&=-r_3 \dot{r}_2 K_x,\notag\\
		h_z&=\dot{r}_2K_y,
	\end{align}	
	then the asymptotic expression of QFI about the parameter $\kappa$ can be obtained as
	\begin{align}
		F_{\kappa} \simeq \frac{64}{9}\delta^4 t^6\Delta ^2 [K_y - K_z]_{|\psi_0\rangle}, \quad \kappa\rightarrow\delta.\label{FK1}
	\end{align}
	For the critical point of $\kappa = \delta$, we have $h_x=0$, and the QFI becomes
	\begin{align}
		F_\kappa=&16(\delta^2 \Delta^2[K_x]_{|\psi_0\rangle} t^4 +\Delta^2[K_y]_{|\psi_0\rangle} t^2
		\notag\\ &+2 \delta \mathrm{Cov}[K_x,K_y]_{|\psi_0\rangle} t^3).\label{FK2}
	\end{align} 	
	One can also choose $\delta$ as the parameter to be estimated, then one has
	\begin{align}
		h_x&=  r_2 \dot{r}_3 (r_3 K_y+r_2K_z), \nonumber\\
		h_y&= r_2 \dot{r}_3 K_x,\nonumber\\
		h_z&= \dot{r}_3K_z.
	\end{align}	    
	In this situation, Eq.~(\ref{FK1}) becomes
	\begin{align}
		F_{\delta} \simeq \frac{64}{9}\kappa^4 t^6 \Delta ^2 [K_z - K_y]_{|\psi_0\rangle}, \quad \delta\rightarrow\kappa,
	\end{align}
	and Eq.~(\ref{FK2}) becomes
	\begin{align}
		F_\delta=&16(\kappa^2 \Delta^2[K_x]_{|\psi_0\rangle} t^4 +\Delta^2[K_z]_{|\psi_0\rangle} t^2
		\notag\\ &+2 \kappa \mathrm{Cov}[K_x,K_z]_{|\psi_0\rangle} t^3).
	\end{align} 
	
	\bibliographystyle{apsrev4-2}	
	\bibliography{ManuscriptRefer}
\end{document}